\documentclass[aps,pre,twocolumn,preprintnumbers,amsmath,amssymb,footinbib]{revtex4-1}
\usepackage{graphicx,epsfig} 
\usepackage{dcolumn} 
\usepackage{wrapfig}
\usepackage{bm} 
\usepackage{color} 
\usepackage[normalem]{ulem} 

\newcommand{\FourierKerMix}[0]{Note that the Fourier transform of the function ${\frac{ab}{a+b} exp(-\frac{1}{2} (a+b)|x| - \frac{1}{2} (a-b) x)}$ is ${1/(\sqrt{2\pi}(1+ ik (b-a)/ab + k^2/ab))}$}

\begin{document}
\title{Dynamics of reaction-diffusion patterns controlled by asymmetric nonlocal coupling as limiting case of differential advection}
\date{\today}

\author{Julien Siebert$^{1}$, Sergio Alonso$^{2}$, Markus B\"ar$^{2}$ and Eckehard Sch\"oll$^{1}$}

\affiliation { 
$^{1}$ Technische Universit\"at, Institut f\"ur Theoretische Physik, Hardenberstr. 36, 10623 Berlin, Germany \\
$^{2}$ Physikalisch-Technische Bundesanstalt, Abbestrasse 2-12, 10587 Berlin, Germany}

\begin{abstract}
A one-component bistable reaction-diffusion system with asymmetric nonlocal coupling is derived as limiting case of a two-component activator-inhibitor reaction-diffusion model with differential advection. 
The effects of asymmetric nonlocal couplings in such a bistable reaction-diffusion system are then compared to the previously studied case of a system with symmetric nonlocal coupling. 
We carry out a linear stability analysis of the spatially homogeneous steady states of the model and numerical simulations of the model to show how the asymmetric nonlocal coupling controls and alters the steady states and the front dynamics in the system. 
In a second step, a third fast reaction-diffusion equation is included which induces the formation of more complex patterns. 
A linear stability analysis predicts traveling waves for asymmetric nonlocal coupling in contrast to a stationary  Turing patterns for a system with symmetric nonlocal coupling. 
These findings are verified by direct numerical integration of the full equations with nonlocal coupling.
\end{abstract}

\maketitle
\section{Introduction}
Pattern formation arises in various chemical and biological systems \cite{HAK83,MIK94,KAP95a,KEE98}. It often appears by the combination of nonlinear chemical reactions and local diffusion. Both stationary and oscillatory patterns have been found in chemical reaction-diffusion systems. Stationary spatially periodic states, {\em i.e.}, Turing patterns, have been observed, for instance, in the chloride-ionide-malonic acid reaction \cite{KAP95a}. Turing patterns are often related to morphogenesis, {\em e.g.}, as a cause for the disposition for stripes in some skin patterns of fish \cite{KON10}. Spiral waves have been studied in the Belousov-Zhabotinsky reaction \cite{KAP95a}. Moreover, spirals were implicated in dangerous arrhythmias in the heart, see e.g. \cite{WIN87,DAV92a}, as well as in spreading depression that plays a role in migraine \cite{GOR83,DAH97}.

The dynamics of reaction-diffusion systems can be affected and controlled by spatial interactions. Global feedback (a closed-loop control, as opposed to open-loop control or external forcing), has been shown to generate a variety of patterns, including cluster states in oscillatory active media \cite{KUR84}. Experimental and theoretical studies in the CO oxidation on platinum surfaces \cite{FAL94,BAE94,FAL95a,BER03,BET03,BET04a} and other catalytic processes \cite{MID92a,SHE96a}, as well as in electro-chemistry \cite{PLE01} or in semiconductors \cite{MEI00b,SCH01} have shown that global feedback can be employed to control propagating waves and to generate spatially periodic patterns such as Turing patterns or travelling waves (for a review see \cite{MIK06}). 

In most of the above cases, the spatial coupling is provided by diffusion and hence possesses reflection symmetry in space. Reflection symmetry is broken by the presence of advective processes.Advection is present  in reaction-diffusion-advection (RDA) systems. RDA systems have been intensively explored as models for heterogeneous catalysis with in- and outflow of the chemical species \cite{MID92a,SHE96b,YOC09}. Their generic features have been subject to detailed mathematical analysis recently 
\cite{YOC09a,YOC10,YOC10a}. Particularly interesting are reaction-diffusion systems with differential flows of the chemical species that can cause novel spatial instabilities known as differential flow-induced chemical instabilities (DIFICIs) \cite{ROV92a,ROV93,BAL95,BAL98}.
Differential flows can be caused by the interaction of one chemical species bound to a catalytic surface (and therefore subject  only to diffusion) with another chemical species in the gas phase above the catalyst. Since the gas phase is typically also used to provide the reactants that adsorb on the surface as well as to remove reaction products, most chemical reactors are operated with a constant flow through the system. 
Therefore, differential advection models have been studied in some detail to model the behavior of flow reactors in chemical engineering \cite{SAT98,TOT99}. 
Beyond this initial application, differential flow appears in a wide variety of systems ,e.g., in marine biology \cite{MAL96} and in the formation of vegetation patterns \cite{KLA99,HAR01,BOR09}.

To obtain formation of periodic patterns in simple reaction-diffusion systems, one requires at least two variables representing one activator with autocatalytic properties and one inhibitor providing a feedback on the activator \cite{PIS06}.
For fast inhibitor dynamics, Petrich et al. \cite{PET94,PIS06} have shown that the well-known FitzHugh-Nagumo model (as well as any two-variable model with linear inhibitor dynamics) reduces to a one-component equation with symmetric nonlocal coupling.
Similarly, three-variable models that include one fast variable with linear dynamics reduce to two-variable models with one or more symmetric nonlocal coupling terms, see e. g. \cite{NIC02,TAN03,SHI04}. Here, we extend this treatment to two- and three-variable models with differential advection. As a result we obtain one- and two-variable models with asymmetric nonlocal coupling.

Systems with nonlocal and global coupling have been already studied for a number of applications. 
Global coupling results when the fast linear variable that is eventually eliminated adiabatically diffuses much faster than the other involved species.
n neuronal systems, especially in the visual cortex, several experimental studies have focused on the nonlocal connectivity, and it is still not clear whether it has a symmetric or asymmetric organization \cite{DAS99,KAN02,XU04b}.
It is often assumed that interacting cortical neurons are coupled in a Mexican-hat fashion, where neighbouring cells excite each other, while distant cells have long-range inhibitory interactions \cite{KAN02}.
In electrochemistry, models with explicit nonlocal coupling have been derived from more elementary models by applying a Green's function formalism, see e.g. \cite{CHR02a}.

It has been observed that nonlocal coupling can induce a variety of spatio-temporal patterns \cite{SHE97a,NIC02,BOR06,COL13,COL14}, basically because it modifies the existence and stability of the homogeneous steady states of the nonlinear system. Such effects have been demonstrated in electro-chemistry \cite{MAZ97}, in the Belousov-Zhabotinsky reaction \cite{HIL01,NIC06} and magnetic fluids \cite{FRI03}, and may induce, for example, fingering \cite{PET94}, remote wave triggering \cite{CHR99}, Turing structures \cite{LI01a}, wave \cite{NIC06} instabilities or spatiotemporal chaos \cite{VAR05} in chemical systems.

Nonlocal coupling schemes that do not change the homogeneous steady states can nevertheless generate patterns \cite{BEL81} or control propagation of fronts and pulses \cite{DAH08,SCH09c,DAH09a} . Nonlocal coupling plays also an essential role in the formation of chimera states in discrete arrays of oscillators \cite{KUR02a,ABR04,HAG12,TIN12,MAR13,OME13}.
It is also known that reaction-diffusion patterns can also be controlled by time-delayed feedback \cite{KIM01,KYR09}.

Here, we consider for the first time systematically the effects of asymmetric nonlocal coupling upon front propagation in a bistable reaction-diffusion system resulting as a limiting case of a two-variable reaction-diffusion system with differential advection. 
Furthermore, we derive a model with one asymmetric and one symmetric nonlocal coupling term from a three variable reaction-diffusion equation with differential advection and analyze the resulting the spatio-temporal dynamics. 
It is shown that asymmetric nonlocal coupling induces a wave instability, while symmetric nonlocal coupling produces Turing patterns in a bistable one-component reaction-diffusion equation. 

The manuscript is organized as follows.
In section \ref{sect.model}, a simple bistable model is introduced.
In section \ref{sect.nlc}, a nonlocal term is added to the system by introducing a fast inhibitor subject to both advection and diffusion. 
In section \ref{sect.2nlc}, a second nonlocal term (derived from a second fast inhibitor which is only diffusive) is added in order to obtain, for a certain choice of parameters, a nonlocal coupling that leaves the steady states of the model unaffected ("non-invasive coupling").
In section \ref{sect.limit}, we explore how this "non-invasive" nonlocal coupling can generate wave instabilities and lead to new patterns.
Influence of the nonlocal coupling term is explored, for all cases, through stability analysis of the homogeneous steady state and numerical simulations.

\section{Model}\label{sect.model}
A one-variable generic bistable reaction-diffusion model \cite{MAL85} in one spatial dimension is considered:
\begin{eqnarray}
\partial_t u &=&  F(u)  + \partial^2_x u,  \label{Eq.schlogl}
\end{eqnarray}
where $u(x,t)$ is the dynamic variable, $F(u)$ is a nonlinear function and the diffusion coefficient is set equal to unity. In particular we use the Schl\"ogl model \cite{SCH72,SCH83a} for the function 
\begin{eqnarray}
 F(u)=-u(u-\alpha)(u-1), \label{Eq.schlogl.F}
\end{eqnarray}
with $0<\alpha<1$. Eq.(\ref{Eq.schlogl}) possesses two stable spatially homogeneous solutions at $u^*_o=0$ and $u^*_+=1$ and an unstable homogeneous solution at $u^*_-=\alpha$. It has a traveling front solution with a monotonic profile with $u(-\infty,t)=1$ and $u(+\infty,t)=0$: 
\begin{eqnarray}
 u(x,t) = \dfrac{1}{2}\left[ 1 - tanh\left( \dfrac{x-ct}{2\sqrt{2}}\right) \right], \label{Eq.schlogl.front}
\end{eqnarray}
The propagation velocity of this front is given by:
\begin{eqnarray}
 c=\sqrt{\dfrac{1}{2}} (1-2 \alpha). \label{Eq.schlogl.velocity}
\end{eqnarray}

\begin{figure}[t]
 \centering
 \includegraphics[width=0.25\textwidth]{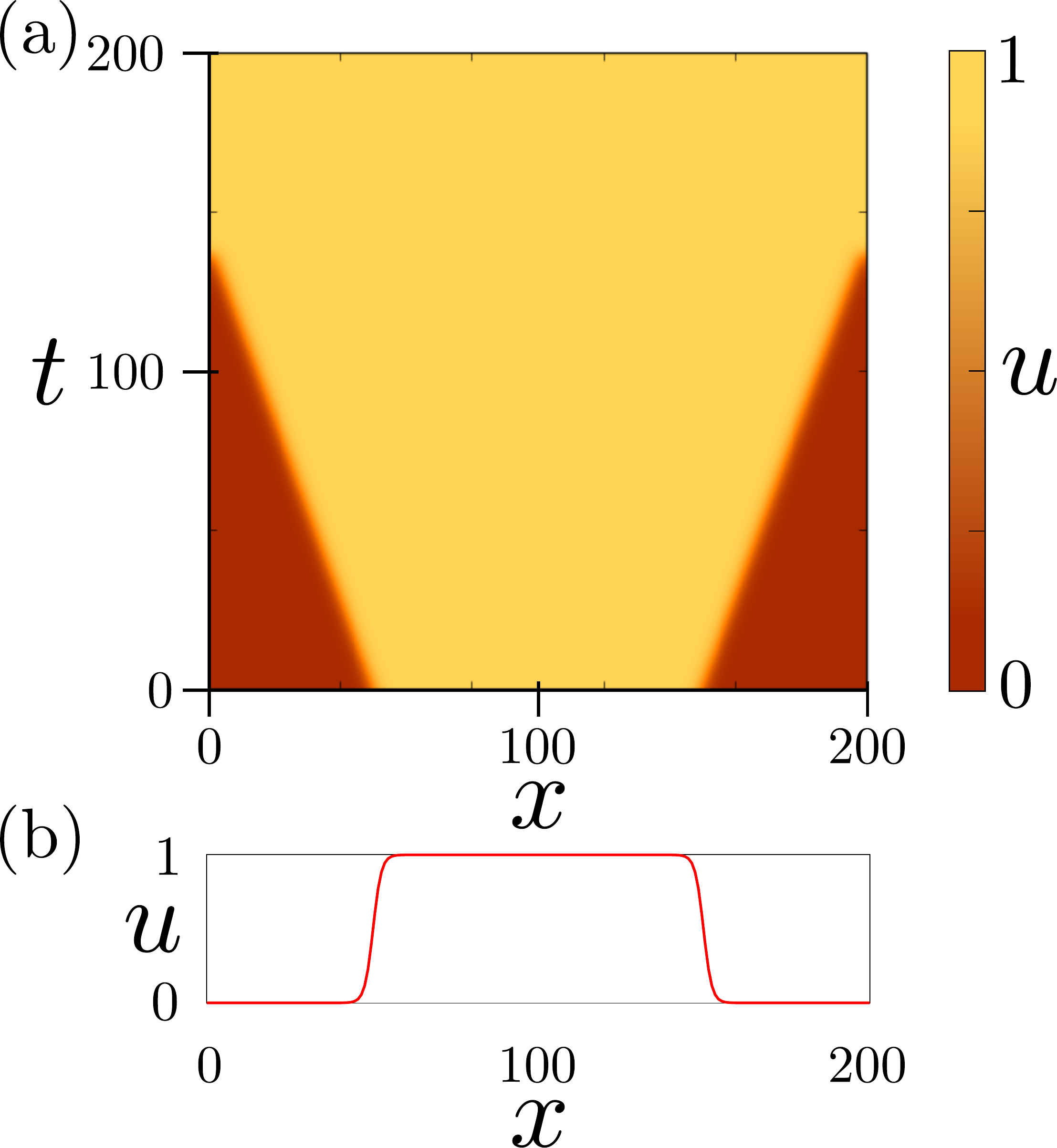}
 \caption{(a) Space-time diagram for the model without control. (b) Initial conditions. Note that periodic boundary conditions are used. Space length: $L = 200$; spatial discretization step: $\Delta x = 0.2$; maximum simulation time $T = 200$; time discretization step: $\Delta t = 0.01$. Parameter $\alpha = 0.25$.}
 \label{fig_stdiag}
\end{figure}

Fig.\ref{fig_stdiag} shows the space-time diagram on a finite domain where a symmetric two-fronts profile was used as initial condition. From these initial conditions, two fronts propagate in opposite directions with the same velocity. For $\alpha < 0.5$, the fronts propagate from the domain of the globally stable state $u^*_+=1$ into the domain of the metastable state $u^*_o=0$, as shown in the  numerical simulation in Fig.\ref{fig_stdiag}(a), and for $\alpha > 0.5$ (note that the system is invariant with respect to the transformation $\alpha \to 1-\alpha, u \to 1-u$), from the globally stable state $u^*_o=0$ to the metastable state $u^*_+=1$.

\section{Nonlocal coupling}\label{sect.nlc}

\subsection{Reaction-diffusion-advection equations}

We now extend Eq.(\ref{Eq.schlogl}) by considering the following reaction-diffusion-advection system:
\begin{eqnarray}
\partial_t u &=&  F(u)  - g w  + \partial^2_x u,  \label{Eq.u} \\  
\tau \partial_t w &=&  h u - f w + \xi \partial_x w+ D \partial^2_x w , \label{Eq.w}
\end{eqnarray}
where all the terms are linear except for the function $F(u)$ that is given by Eq.(\ref{Eq.schlogl.F}) above. 
The two concentrations $u$ and $w$ correspond to the activator and the inhibitor, respectively, and are linearly coupled by the terms $-gw$ and $hu$ with real constants $g$ and $h$. The constant $f$ will be set to be $f=1$ in the following, $\tau$ is the inhibitor relaxation time and $D$ corresponds to the inhibitor diffusion coefficient ($D>0$). The coefficient $\xi$ corresponds to the advection velocity, and depending on the direction of the advective flow, can be either positive or negative.

\begin{figure}[t]
\centering
  \resizebox{0.4\textwidth}{!}{
\includegraphics[]{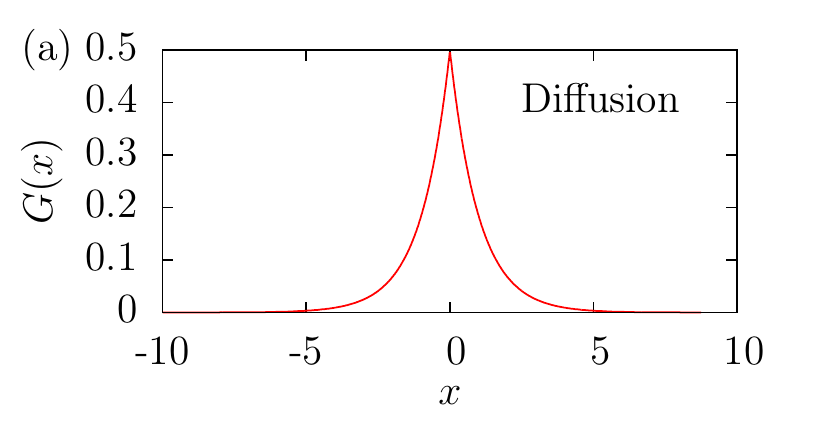}
  }\\
  \vspace{-0.5cm}
  \resizebox{0.4\textwidth}{!}{
\includegraphics[]{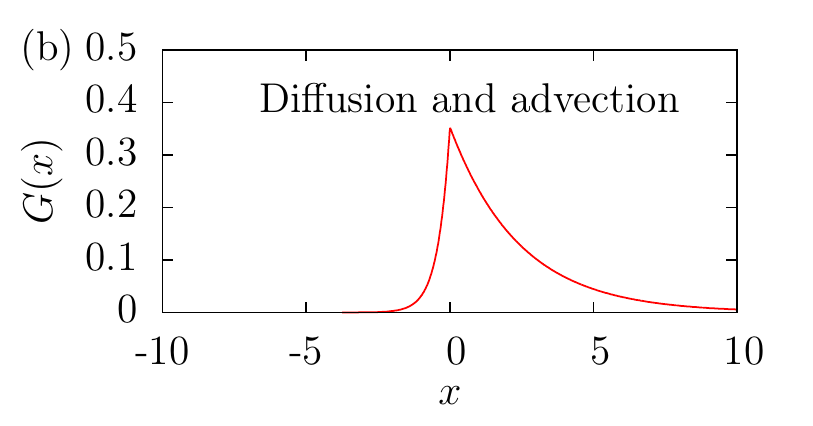}
  }\\
    \vspace{-0.5cm}
  \resizebox{0.4\textwidth}{!}{
\includegraphics[]{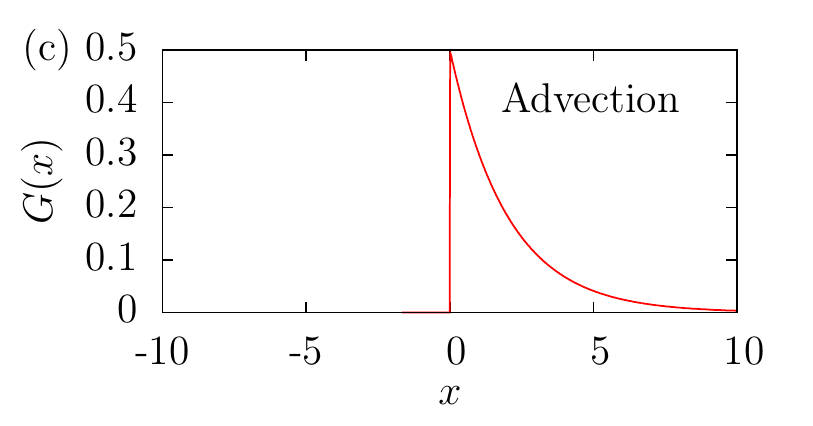}
  }\\

\caption{Kernel functions for three different cases: (a) pure diffusion with $D=1$ and $\xi=0$, (b) diffusion and advection with $D=1$ and $\xi=2$, (c) pure advection with $D=0$ and $\xi=2$ (c). 
The other parameter is $f=1$. } 
\label{figker}
\end{figure}

We assume that $\tau \ll 1$ is small - which corresponds to the case of fast inhibitor dynamics $w$.
In the limit $\tau \rightarrow 0$, we can eliminate the variable $w$ in Eq.(\ref{Eq.u}) by solving the linear Eq.(\ref{Eq.w}). 
First, we apply the Fourier transform to the inhibitor equation for the case $\tau = 0$:
\begin{eqnarray}
h \hat{u}(k) - (f + i k \xi + D k^2)\hat{w}(k)  = 0,  \label{Eq.fourier} 
\end{eqnarray}
where $\hat{u}(k)$ and $\hat{w}(k)$ are the Fourier transforms of $u$ and $w$, respectively. Solving for $\hat{w}$ and
transforming the resulting expression back to the original variable $x$ yields:
\begin{eqnarray}
w(x,t)= \frac{1}{\sqrt{2 \pi}} \int_{-\infty}^{\infty} \mathrm{d}k \, e^{ikx} \frac{h  \hat{u}(k)}{f + i k \xi + D k^2}.  \label{Eq.fourier_1} 
\end{eqnarray}

The integral over $k$ is the product of two Fourier transforms \footnote{\FourierKerMix{}}:
\begin{eqnarray}
w(x,t)= \frac{h}{\sqrt{2 \pi}} \int_{-\infty}^{\infty} \mathrm{d}k \, e^{ikx} \sqrt{2\pi}\hat{u}(k)\hat{G}(k), \label{Eq.fourier_2} 
\end{eqnarray}
and, by definition, can be written as a convolution integral:
\begin{eqnarray}
 w(x,t) = h \int_{-\infty}^{\infty} \mathrm{d}x' G(x')u(x-x',t),
\end{eqnarray}
with the nonlocal integration kernel:
\begin{eqnarray}
G(x) \equiv \frac{f}{\sqrt{\xi^2+4Df}} e^{-\frac{\sqrt{\xi^2+4Df}}{2D} |x|} e^{\frac{\xi}{2D} x}.  \label{Eq.ker}
\end{eqnarray}
Hence Eqs.(\ref{Eq.u}) and (\ref{Eq.w}) can be replaced by a reaction-diffusion equation for $u(x,t)$,  for the limit $\tau = 0$  :
\begin{eqnarray}
\partial_t u &=&  F(u)  + \partial^2_x u  - \sigma \int_{-\infty}^{\infty} \mathrm{d}x' G(x') u(x-x'), \label{Eq.u.nlc_2}   
\end{eqnarray}
where the nonlocal coupling strength is defined by ${\sigma = gh}$. The parameter $\sigma$ is positive or negative depending on the particular choice of parameters $g$ and $h$. Note, that Eq.(\ref{Eq.u}), (\ref{Eq.w}) represent an activator-inhibitor model if one chooses $g>0$ and $h>0$. As a result, the choice $\sigma > 0$ corresponds to long-range inhibition in Eq.(\ref{Eq.u.nlc_2}).

From Eq.(\ref{Eq.ker}), we can recover analytically the limiting cases of pure diffusion and pure advection in the fast inhibitor equation. For example, in the limit $\xi \rightarrow 0$, one obtains the kernel for pure diffusion (symmetric kernel):
\begin{eqnarray}
G(x) = \dfrac{1}{2}\sqrt{\dfrac{f}{D}} \exp \left(-\sqrt{\frac{f}{D}} |x| \right), \label{Eq.diff} 
\end{eqnarray}
which is illustrated in Fig.\ref{figker}(a). On the other hand, in the limit $D \rightarrow 0$, $\xi>0$ and after a Taylor expansion, one recovers the kernel for pure advection:
\begin{eqnarray}
G(x) = \dfrac{f}{\xi} \exp \left(- \dfrac{f}{\xi} x \right) \Theta (x), \label{Eq.adv} 
\end{eqnarray}
where $\Theta (x)$ is the Heaviside function. An example is shown in Fig.\ref{figker}(c). The general case (asymmetric kernel) is depicted in Fig.\ref{figker}(b). Note that the kernel is normalized:
\begin{eqnarray}
 \int_{-\infty}^{\infty} \mathrm{d}x \, G(x) = 1. \label{eq.ker.norm}
\end{eqnarray}

\begin{figure}[t]
  \resizebox{0.4\textwidth}{!}{
	\includegraphics[]{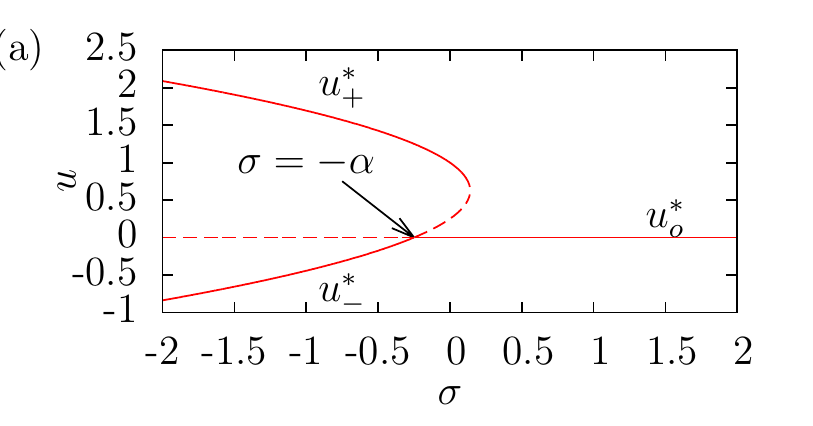}
  }\\
  \resizebox{0.4\textwidth}{!}{
	\includegraphics[]{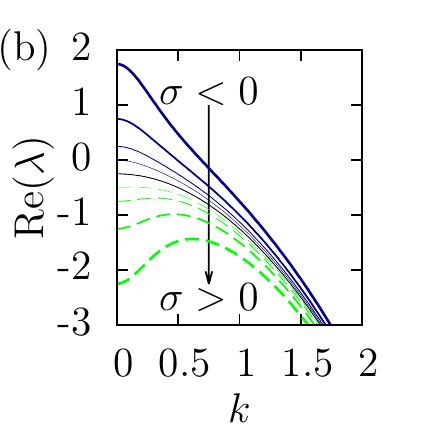}
    \hspace{-1cm}
	\includegraphics[]{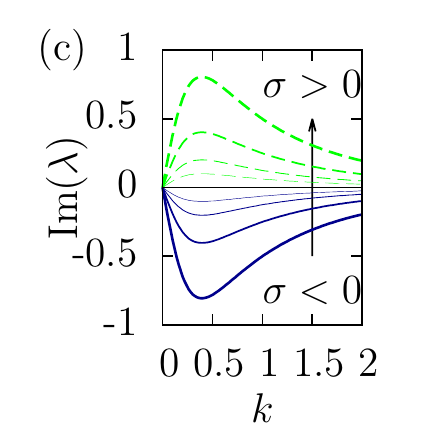}
  }

\caption{
  (a) Spatially homogeneous solutions of Eq.(\ref{Eq.u.nlc_2}) for different values of nonlocal coupling strength $\sigma$.
  Solid and dashed lines represent stable and unstable solutions, respectively.
  (b), (c) Dispersion relations Re $\lambda(k)$, Im $\lambda(k)$ of the solution $u^*_o$ for different values of $\sigma$. 
  Arrows show the direction of increasing values of $\sigma$ (solid blue: $\sigma < 0$. dashed green: $\sigma > 0$).
  Parameters: $\alpha=0.25$, $D=1$, $\xi=2$, $f=1$, $\sigma=-1, -0.5, -0.25, -0.1, 0, 0.1, 0.25, 0.5, 1$.
}
\label{fighom}
\end{figure}

\subsection{Stability analysis of the homogeneous states}

Equation (\ref{Eq.u.nlc_2}) corresponds to a reaction-diffusion equation with nonlocal coupling.
This equation, with the choice $F(u)=-u(u-\alpha)(u-1)$ for the nonlinear reaction term, has three spatially homogeneous steady states: $u^*_o=0$ and $u^*_{\pm} =\left( (1+\alpha) \pm \sqrt{(1+\alpha)^2 - 4(\alpha + \sigma)}\right)/2$, for $\sigma \leq (1-\alpha)^2/4$. 
The dependence of the solutions on the parameter $\sigma$ is depicted in Fig.\ref{fighom}(a). 
The nonlocal coupling modifies the homogeneous solution of the system, and even two of the solutions cross over for $\sigma=-\alpha$. 

In order to obtain the dispersion relation, we perform a linear stability analysis around the solution $u^*_o=0$ by setting $u(x,t) = u^*_o + \delta u$ in equation (\ref{Eq.u.nlc_2}) with $\delta u =\delta u_o e^{ikx}e^{\lambda t}$ and small $\delta u_o$. This leads to:
\begin{eqnarray}
\lambda &=& F'(u^*_o)  - k^2  - \sigma \left(\frac{1}{f + i\xi k + D k^2}\right). \label{Eq.DisRel}
\end{eqnarray}
The real and imaginary parts are shown in Fig.\ref{fighom}(b) and Fig.\ref{fighom}(c), respectively, for a range of values of $\sigma$. 
For large negative nonlocal coupling strength $\sigma < - \alpha$ the solution $u^*_o=0$ becomes unstable.
For large positive coupling strength $\sigma >  (1-\alpha)^2/4$, $u^*_o=0$ is the only real solution. 
The dispersion relation for large enough positive values of the coupling strength $\sigma$ shows a maximum at finite wave number $k > 0$, see Fig.\ref{fighom}(b).
The conditions to have such a maximum at a finite wave number is $\sqrt{\sigma D} > f$ and $\sigma > 0$ (long-range inhibition). This does not necessarily lead to a Turing instability, which would require that $f > 2 \sqrt{\sigma D} +\alpha D$. It is easy to see that this condition  and the condition for a maximum of the dispersion curve at $k \ne 0$ cannot be fulfilled as long as the parameters $\alpha$ and $D$ are positive. The choice $\alpha > 0$ is necessary since the Schl\"ogl model requires $0 < \alpha < 1$ (see  section II). 
For $\xi \neq 0$, the dispersion relation is complex, and exhibits an oscillatory instability for $\sigma < -\alpha$.

\subsection{Velocity of the fronts}
\label{velocity_analytics}
The previous linear stability analysis of the homogeneous steady states shows that the system still exhibits bistability for $\sigma< (1-\alpha)^2/4$. 
Hence, while the front solution may disappear for sufficiently large positive values of $\sigma$, for $\sigma < (1-\alpha)^2/4$ propagating fronts between the two stable homogeneous steady states can still be expected, albeit with a modified propagation velocity.
In this section, we derive an analytical approximation for the velocity of the fronts under nonlocal coupling.

For $|\sigma|$ not too large, we may assume that the front still propagates with a stationary profile which is not essentially distorted, and with a constant propagation velocity $c$, and we can transform Eq.(\ref{Eq.u.nlc_2}) to the co-moving frame $\zeta = x - ct$:

\begin{eqnarray}
 0 = F(u) + c u' + u'' - \sigma\int_{-\infty}^{\infty} \mathrm{d}\zeta' G(\zeta')u(\zeta-\zeta'), \label{Eq.velo.comov}
\end{eqnarray}
with $u' = \partial_{\zeta}u(\zeta)$.
Assuming a monotonic front profile with $u(\zeta=-\infty)=1$, $u(\zeta=+\infty)=0$, like the right front in Fig.\ref{fig_stdiag}(b), we can determine the front velocity without explicitly solving Eq.~(\ref{Eq.velo.comov}) by the following argument, cf. \cite{SCH01}, Ch.3.1.

Multiplying Eq.(\ref{Eq.velo.comov}) by $u'(\zeta)$ and integrating over $\zeta$ from $-\infty$ to $+\infty$ yields
\begin{eqnarray}
 0 &=& \int_1^0 F(u)\mathrm{d}u +  c\int_{-\infty}^{+\infty} (u')^2  \mathrm{d}\zeta + \sigma G_+,
\end{eqnarray}
with
\begin{eqnarray}
G_+ &=& -\int_{-\infty}^{+\infty} \mathrm{d}\zeta' G(\zeta') \int_{-\infty}^{+\infty}\mathrm{d}\zeta u(\zeta-\zeta')u'(\zeta),  
 \label{Eq.velo.int}
\end{eqnarray}
where the boundary conditions $u'(\zeta=-\infty)=u'(\zeta=+\infty)=0$ have been used to eliminate the diffusion term $\int_{-\infty}^{+\infty} \mathrm{d}\zeta u''(\zeta) u'(\zeta) = \left.\frac{1}{2} u'(\zeta)^2 \right]_{-\infty}^{+\infty}$.

The nonlocal coupling term can be approximately evaluated by using the fact that the front profile is close to a Heaviside function $u(\zeta) \approx u_+^*\Theta(-\zeta)$, and $u'(\zeta) \approx -\delta(\zeta)$ is close to Dirac's delta function:
\begin{eqnarray}
G_+\approx \int_{-\infty}^{+\infty} \mathrm{d}\zeta' G(\zeta') u(-\zeta') \approx u_+^*\int_{0}^{\infty} \mathrm{d}\zeta' G(\zeta'), \label{Eq.velo.int.coupling}
\end{eqnarray}
which is always positive, and for a symmetric coupling kernel (pure diffusion)
the integral is equal to $1/2$, while for the limit case of pure advection it is equal to $1$, see Fig.~\ref{figker}.

The front propagation velocity then follows from Eq.(\ref{Eq.velo.int}):
\begin{eqnarray}
 c=\dfrac{ \int_0^1 F(u)\mathrm{d}u - \sigma G_+ }{ \int_{-\infty}^{+\infty} (u')^2\mathrm{d}\zeta}. \label{Eq.c.alpha<0.5}
\end{eqnarray}

Since $\int_{-\infty}^{+\infty} (u')^2  \mathrm{d}\zeta >0$, and $\int_0^1 F(u)\mathrm{d}u = (1-2\alpha)/12>0$ for $\alpha<0.5$, the propagation velocity $c$  is positive for $\sigma=0$ and remains so with nonlocal coupling provided that $\sigma<0$ or $|\sigma|$ is not too large in case of $\sigma>0$.

For $\sigma = 0$ we recover the front velocity of the Schl\"ogl model without feedback (see Eq.(\ref{Eq.schlogl.velocity})) where $\int_{-\infty}^{+\infty} (u')^2  \mathrm{d}\zeta = 1/(2\sqrt{2})$ can be evaluated explicitly using the front profile given by Eq.(\ref{Eq.schlogl.front}). The nonlocal coupling accelerates or decelerates the front for $\sigma < 0$ or $\sigma > 0$, respectively.
The detailed dependence upon the parameters $\sigma, D, \xi, f$ can be explored by substituting the homogeneous steady state $u^*_+ =\left( (1+\alpha) + \sqrt{(1+\alpha)^2 - 4(\alpha + \sigma)}\right)/2$, and evaluating the integral Eq.~(\ref{Eq.velo.int.coupling}).

Note that the same reasoning can be applied to the left front profile in Fig.\ref{fig_stdiag}(b) with $u(-\infty,t)=0$, $u(+\infty,t)=1$. In this case, the velocity is:
\begin{eqnarray}
 c=\dfrac{ -\int_0^1 F(u)\mathrm{d}u + \sigma G_- }{ \int_{-\infty}^{+\infty} (u')^2 \zeta} < 0, \label{Eq.c.alpha>0.5}
\end{eqnarray}
with $G_-\approx \int_{-\infty}^{+\infty} \mathrm{d}\zeta' G(\zeta') u(-\zeta') \approx u_+^*\int_{-\infty}^{0} \mathrm{d}\zeta' G(\zeta')$,  
which is always positive, and for a symmetric coupling kernel (pure diffusion)
the integral is equal to $1/2$, while for the limit case of pure advection it is equal to $0$, see Fig.~\ref{figker}.  It is interesting to note that the asymmetry of the kernel leads to an asymmetry in the propagation velocities of fronts to the left and to the right. Only with pure diffusion we obtain symmetric modifications of the propagation velocity by nonlocal coupling. Generally, for $\xi>0$, the modification of the front propagating to the left is weaker than of the front propagating to the right, and vanishes in the limit of pure advection.

\subsection{Numerical simulations}

We have performed numerical simulations of Eq.(\ref{Eq.u.nlc_2}) for different values of the parameters of the nonlocal coupling. In the limit $\sigma = 0 $ we recover Eq.(\ref{Eq.schlogl}) and the travelling solution shown in Fig.\ref{fig_stdiag}. Note that in Fig.\ref{fig_stdiag} two fronts are generated by the initial conditions, one front propagating to the left, the other to the right. For $\sigma < 0$ ($\sigma >0$) the velocity of the fronts increases (decreases).

\begin{figure}[t]
 \centering
 \includegraphics[width=0.4\textwidth]{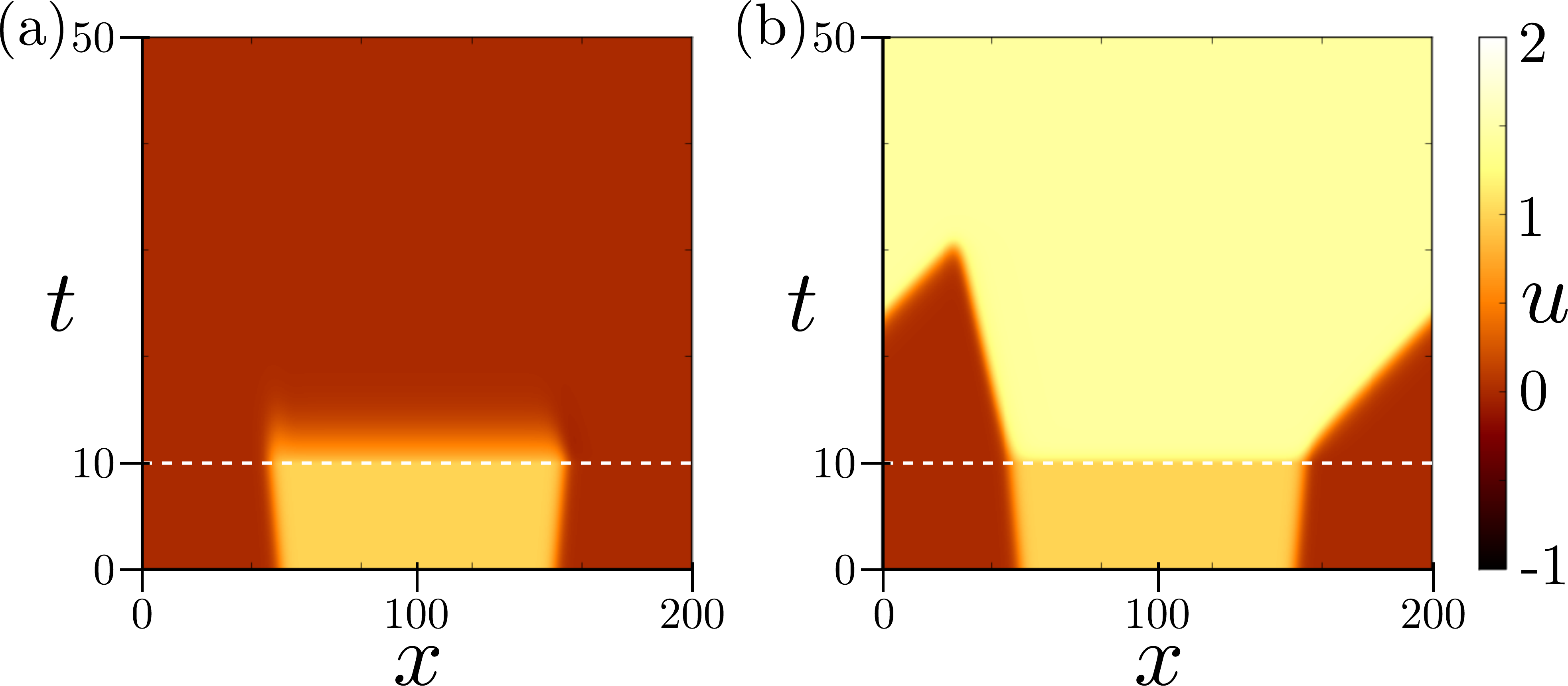}
 \caption{Space-time diagram (numerical simulation of Eq.(\ref{Eq.u.nlc_2}) with parameters $\alpha = 0.25$, $f=1$, $\xi=2$, $D=1$). Control is applied at time $t = 10$ (white dashed line). (a) Front suppression ($\sigma = 0.2$). (b) Modification of the front velocity ($\sigma = -0.5$). All other simulation parameters as in Fig.\ref{fig_stdiag}.}
 \label{stdiag.accel.suppr}
\end{figure}

In Fig.~\ref{stdiag.accel.suppr} two space-time plots are shown for positive
(a) and negative (b) $\sigma$. In Fig.~\ref{stdiag.accel.suppr}(a), the upper homogeneous steady state $u_+^*$ does not exist anymore in the controlled system, and fronts are no longer possible. The only remaining homogeneous steady state $u_0^*=0$ fills the whole system. In Fig.~\ref{stdiag.accel.suppr}(b), both fronts are accelerated, but the velocity change of the right front is larger than that of the left front, as predicted from our analysis in Sect.\ref{velocity_analytics}. Due to the periodic boundary conditions the right front eventually reenters from the left boundary, and the whole system is finally swept into the homogeneous steady state $u_+^*$. Note that $u_+^*$ is increased by the coupling, as shown in the previous section.

\begin{figure}[t]
 \centering
 \resizebox{0.5\textwidth}{!}{
 \hspace{-0.5cm}
\includegraphics[]{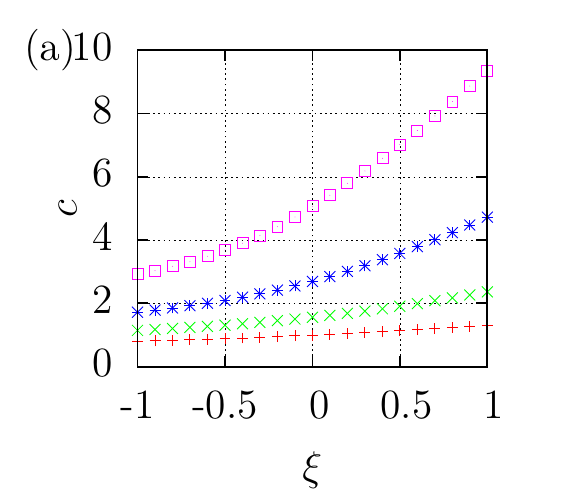}
 \hspace{-1cm}
\includegraphics[]{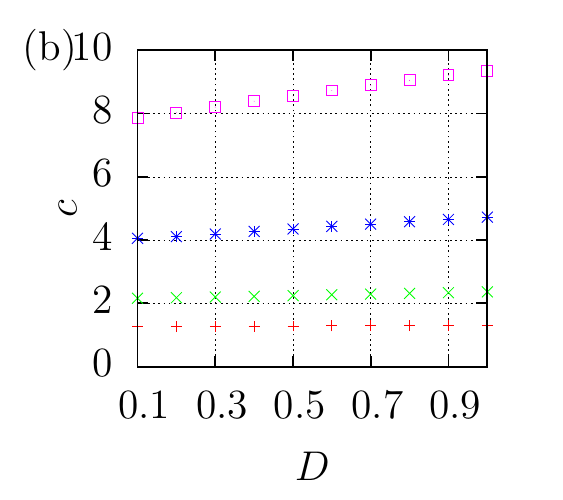}
 }
 \caption{Effect of advection parameter $\xi$ and diffusion constant $D$ on the velocity $v$ of the right front for $\alpha = 0.25$, $f=1$, $\sigma=-0.25 (+),-0.5 (\times),-1 (\ast),-2 (\boxdot)$, (a) $D=1$; (b) $\xi = 1$. Simulation parameters as in Fig.\ref{fig_stdiag}.}
 \label{figvel}
\end{figure}

The parameter $\xi$ breaks the symmetry of the kernel. It increases the velocity of one front while decreasing the velocity of the other. Fig.\ref{figvel}(a) shows the velocity of the right front. The bigger the values of $\sigma$ and $\xi$ are, the larger is the front velocity (this effect is symmetric for the other front).
The increase of the parameter D in the asymmetric nonlocal coupling ($\xi > 0$) produces faster fronts, see Fig.\ref{figvel}(b). The same effect is observed for the symmetric nonlocal coupling ($\xi = 0$) case.

Only front solutions or the corresponding asymptotic homogeneous steady states have been observed in the simulations of Eq.(\ref{Eq.u.nlc_2}), for the parameters chosen here. We do not observe any travelling waves or Turing patterns in this case, in agreement with the predictions of the stability analysis in Section III B.

\section{Two nonlocal couplings}\label{sect.2nlc}

\subsection{Reaction-diffusion-advection equations}

In the next step, a third variable $v$ can be added to the generic reaction-diffusion-advection system considered above in Eqs.(\ref{Eq.u}) and (\ref{Eq.w}):
\begin{eqnarray}
\partial_t u &=&  F(u)  - g w - g_2 v  + \partial^2_x u,  \label{Eq.u2} \\  
\tau \partial_t w &=&  h u - f w + \xi \partial_x w+ D \partial^2_x w , \label{Eq.w2} \\
\tau_2 \partial_t v &=&  h_2 u - f_2 v + D_2 \partial^2_x v , \label{Eq.v2}
\end{eqnarray}
where the second inhibitor $v$ is linearly coupled with the activator $u$ by the terms $-g_2 v$ and $h_2 u$. 
The parameter $D_2$ is the diffusion coefficient of the second inhibitor, $\tau_2$ is its relaxation time, and $f_2$, $h_2$, $g_2$ are constants.

Similarly as in the previous section we assume that $\tau$ and $\tau_2$ are small. 
This corresponds to the case of two fast inhibitors ($w$, $v$). 
Using the kernels from Eq.(\ref{Eq.ker}) and Eq.(\ref{Eq.diff}), an extended reaction-diffusion equation with nonlocal coupling results:

\begin{eqnarray}
\partial_t u =  F(u)  + \partial^2_x u  - \sigma_1 \int_{-\infty}^{\infty} \mathrm{d}x' G(x') u(x-x') \nonumber \\ 
- \sigma_2 \int_{-\infty}^{\infty} \mathrm{d}x' G_2(x') u(x-x'), \label{Eq.u.2nlc}   
\end{eqnarray}
where we define the strength of the second nonlocal coupling $\sigma_2 = g_2 h_2$ and set $\sigma_1 \equiv \sigma$.

\begin{figure}[t]
\centering
  \resizebox{0.4\textwidth}{!}{
    \hspace{-0.35cm}
\includegraphics[]{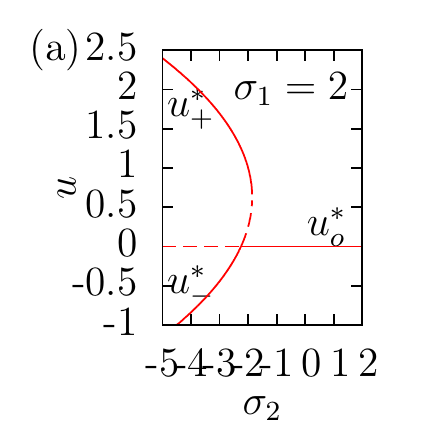}
    \hspace{-0.5cm}
\includegraphics[]{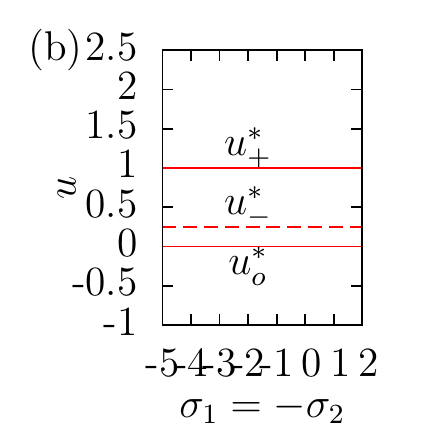}
  }\\
  \resizebox{0.4\textwidth}{!}{
\includegraphics[]{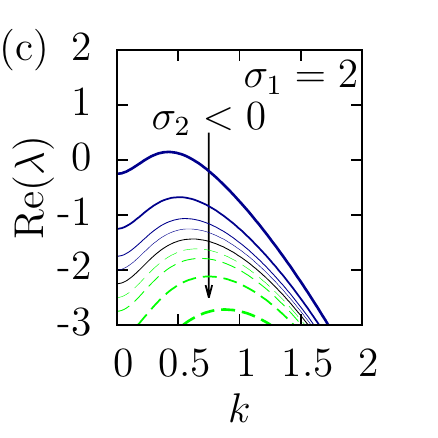}
\includegraphics[]{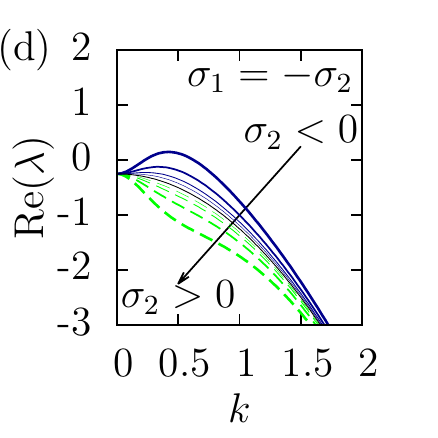}
  }

\caption{Spatially homogeneous solutions of Eq.(\ref{Eq.u.2nlc}) for different values of $\sigma_2$ and (a) ${\sigma_1=2}$; (b)~${\sigma_2=-\sigma_1}$. Solid and dashed lines represent stable and unstable solutions, respectively. (c),(d): Dispersion relation of the solution $u^*_o$ in Eq.(\ref{Eq.DisRel2nlc}) for different values of ${\sigma_2 = -2,-1,-0.5,-0.25,0,0.25,0.5,1,2}$, for (c) $\sigma_1=2$ and (d) $\sigma_2=-\sigma_1$. 
Arrows show the direction of increasing value of $\sigma_2$ (solid blue: $\sigma_2 < 0$; dashed green: $\sigma_2 > 0$). Parameters: $ \alpha=0.25 $, $f=1$, $D=1$, $\xi=2$, $f_2 = 1$, $D_2= 1$.}
\label{fighom2}
\end{figure}

\subsection{Stability analysis of the homogeneous states}

Equation (\ref{Eq.u.2nlc}) has three homogeneous steady state solutions: $u^*_o=0$ and
\begin{eqnarray}
 u^*_{\pm} =\left( (1+\alpha) \pm \sqrt{(1+\alpha)^2 - 4(\alpha + \sigma_1 + \sigma_2)}\right)/2,\nonumber
\end{eqnarray}
for ${\sigma_1 + \sigma_2 \leq (1-\alpha)^2/4}$. 
The stability analysis of the solution $u^*_0$ yields the dispersion relation:
\begin{eqnarray}
\lambda = F'(u^*_0)  - k^2  - \sigma_1 \left( \frac{1}{f + i\xi k + D k^2}\right)  \nonumber \\
- \sigma_2 \left(\frac{1}{f_2 + D_2 k^2} \right). \label{Eq.DisRel2nlc}
\end{eqnarray}

For $\sigma_2=0$ we recover the results obtained in Sect.\ref{sect.nlc}.
The dependence of the solutions on the parameter $\sigma_2$ is shown in Fig.\ref{fighom2}(a),(b). 
For fixed $\sigma_1$, a maximum appears in the dispersion relation. Decreasing $\sigma_2 < 0$ shifts the whole dispersion relation to higher values, see Fig.\ref{fighom2}(c), and for sufficiently large negative $\sigma_2$, unstable complex modes lead to a wave instability. 
For the particular condition $\sigma_2 = - \sigma_1$ the homogeneous solutions are the same as in the Schl\"ogl model without coupling ($u^*_o=0$, $u^*_+=1$ and $u^*_-=\alpha$), see Fig.\ref{fighom2}(b). 
For this condition the combined effect of the two nonlocal couplings on the homogeneous states may hence be called "non-invasive".
However, the stability of the solutions changes due to the nonlocal coupling terms and complex modes grow in time, {\em i.e.} a  wave instability at finite wave number $k \neq 0$ occurs, see Fig.\ref{fighom2}(d). 

\begin{figure}[t]
 \centering
 \includegraphics[width=0.475\textwidth]{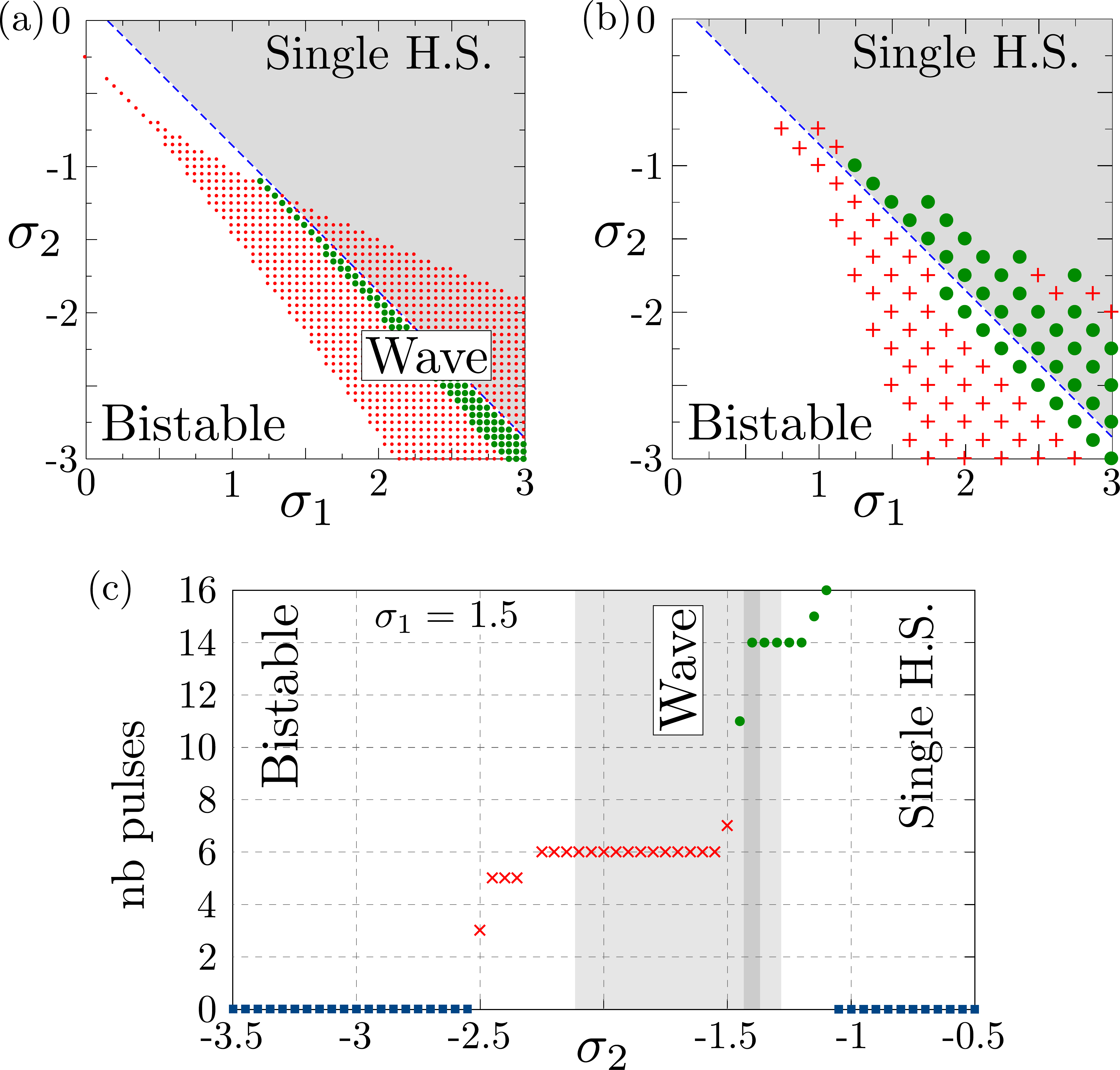}
 \caption{ 
  (a) Stability diagram of the homogeneous steady states (H.S.) numerically obtained from dispersion relation Eq.(\ref{Eq.DisRel2nlc}). 
  Single H.S. (gray shaded): there is only one real spatially homogeneous solution $u^*_o$.
  Bistability (white): two homogeneous solutions are stable.
  Wave instability (light red dots): one of the solutions exhibits a wave instability.
  Strong wave regime (dark green dots): both solutions exhibit a wave instability.\\
  (b) Patterns obtained from simulation of Eq.(\ref{Eq.u.2nlc}).
  Single H.S. (gray shaded): the fronts are suppressed, cf. Fig.\ref{stdiag.accel.suppr}(a).
  Bistability (white): acceleration or deceleration of the fronts (cf. Fig.\ref{stdiag.accel.suppr}(b)). Wave regime (green circles): travelling wave patterns, see Fig.\ref{stdiag2}(c); (red crosses): coexistence of waves and homogeneous states, see Fig.\ref{stdiag2}(c),(d).\\
  (c) Number of pulses observed in the simulation (for fixed $\sigma_1 = 1.5$). Single H.S., bistability and wave region (gray) are shown. Green circles: Travelling waves patterns. Red crosses: Coexistence of travelling wave patterns and homogeneous steady state.
  Parameters $\alpha = 0.25$, $f=1$, $\xi = 2$, $D = 1$, $f_2=1$, $D_2=0.1$. Initial conditions and simulation parameters as in Fig.\ref{fig_stdiag}.
 }
 \label{figphasediag}
\end{figure}

We have also performed a stability analysis of the other spatially homogeneous solutions. The results are plotted in a phase diagram in Fig.\ref{figphasediag}(a). For the homogeneous solutions ($u^*_o$,$u^*_+$), three situations are possible: (i) both solutions are unstable with respect to a wave bifurcation, (ii) one of the solution is stable, while the other is unstable with respect to a wave bifurcation, or (iii) both solutions are stable, see  Fig.\ref{figphasediag}(a).

\subsection{Numerical simulations}
We have numerically integrated Eq.~(\ref{Eq.u.2nlc}) for different values of the parameters. We keep the first kernel asymmetric and we change both amplitudes of the nonlocal couplings $\sigma_1$ and $\sigma_2$. The corresponding phase diagrams in the ($\sigma_1, \sigma_2$) plane are shown in Fig.\ref{figphasediag}(a),(b). Keeping $\sigma_2<0$, we allow both positive and negative values of $\sigma_1$. For negative values of $\sigma_1$ the front is accelerated, while for positive values different behaviors are obtained. Large positive values of $\sigma_1$ decelerate the front, and intermediate values $\sigma_1 \approx - \sigma_2$ generate wave patterns. The regime of wave formation is extended, see Fig.\ref{figphasediag}(a),(b), and depends upon the parameters of the nonlocal coupling.
We have explored the effect of the ratio $D/D_2$ upon the regime of traveling wave patterns both by a stability analysis of the homogeneous solutions ($10^{-8} \leq D/D_2 \leq 10^8$) and by simulation ($1/50 \leq D/D_2 \leq 50$). For decreasing $D/D_2<10$, we have found that traveling waves still exist. However, this regime tends to be smaller than in Fig.\ref{figphasediag}(a),(b), and is also shifted to larger positive values of $\sigma_1$ and larger negative values of $\sigma_2$. 
For increasing $D/D_2$, this regime tends to grow until it reaches an asymptotic limit (the results of the stability analysis of the homogeneous solutions remain qualitatively similar for $D/D_2 > 10^3$).

\begin{figure}[t]
 \centering
 \includegraphics[width=0.4\textwidth]{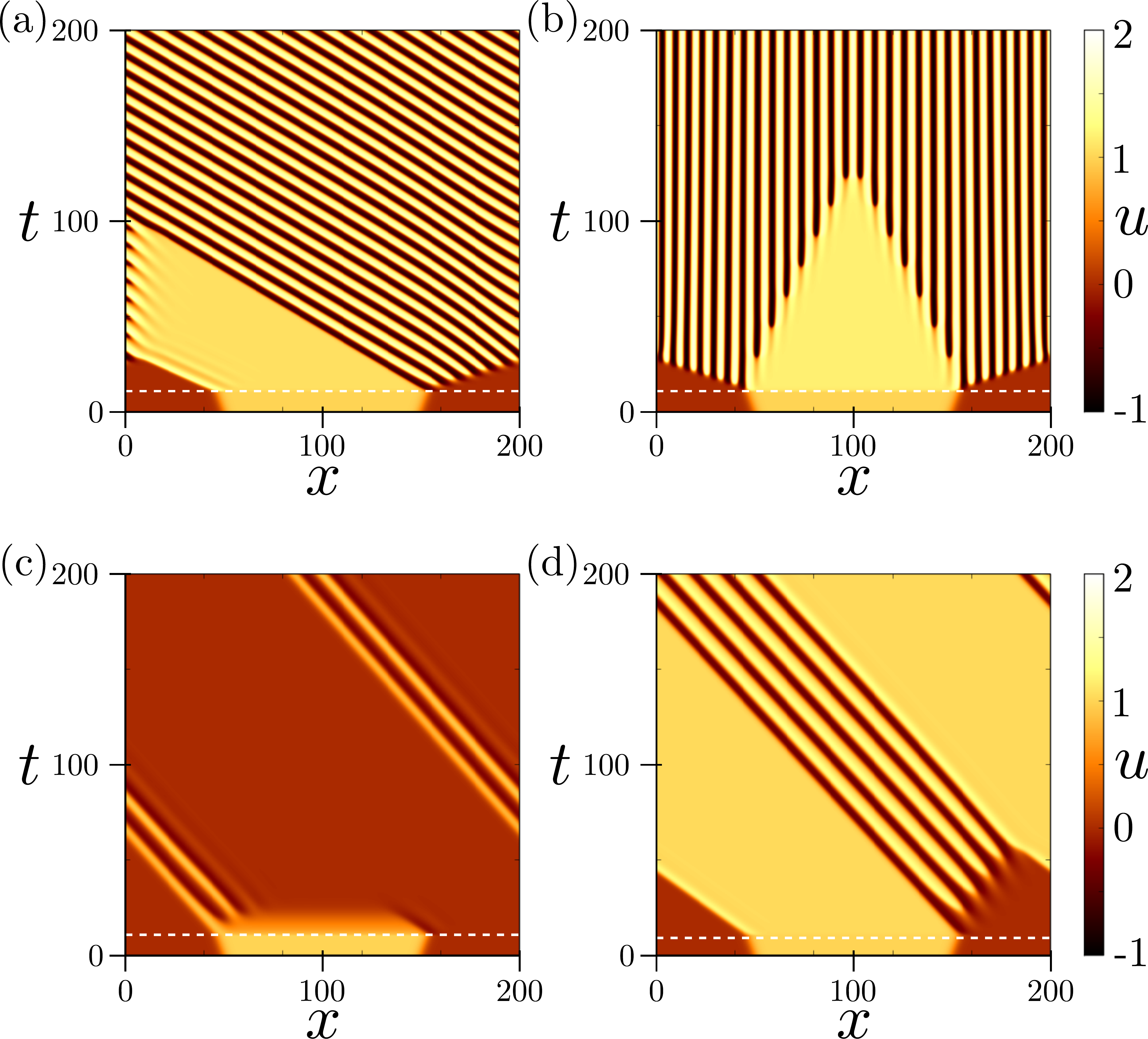}
 \caption{Space-time diagram (numerical simulation of Eq.(\ref{Eq.u.2nlc}) with parameters $\alpha = 0.25$, $f=1$). Control is applied at time $t = 10$ (white dashed line). (a) Travelling wave pattern ($\xi=2$, $D=1$, $D_2=0.1$, $\sigma_1 = -2.0$, $\sigma_2 = 2.0$). (b) Turing pattern ($\xi=0$, $D=0.1$, $D_2=0.1$, $\sigma_2 = -\sigma_1 = 5$). (c) Coexistence of travelling wave patterns and homogeneous steady state $u^*_o$ ($\xi=2$, $D=1$, $D_2=0.1$, $\sigma_1 = -1$ and $\sigma_2=0.75$). (d) Coexistence of travelling wave patterns and homogeneous steady state $u^*_+$ ($\xi=2$, $D=1$, $D_2=0.1$, $\sigma_1 = -1$ and $\sigma_2=1$).}
 \label{stdiag2}
\end{figure}

From the simulation of the differential equation (Fig.\ref{figphasediag}(b),(c)), we observe that the system can exhibit either travelling wave patterns (Fig.\ref{stdiag2}(a)) or coexistence of travelling waves and homogeneous steady state (Fig.\ref{stdiag2}(c),(d)) outside the regime of instability of the homogeneous steady states ($u^*_o$,$u^*_+$).

Note that in the limit of $\xi = 0$ (symmetric kernel $G(x)$ due to diffusion only) Turing patterns are obtained instead of travelling waves (Fig.\ref{stdiag2}(b)).

\section{Limit case $D_2=0$}\label{sect.limit}

\subsection{Resulting nonlocal coupling equation}
In this section, Eq.(\ref{Eq.u.2nlc}) is considered in the particular limit of zero diffusion ($D_2 \rightarrow 0$) of the third variable in Eqs.(\ref{Eq.u2}) and (\ref{Eq.v2}). 
In this case, the third variable $v$ simply follows the variable $u$ adiabatically and we obtain $v=u h_2/f_2$. Defining $\sigma_2= g_2 h_2/f_2$ and imposing $\sigma_1 = - \sigma_2 \equiv \sigma$, we finally obtain a nonlocal coupling in a form which vanishes for homogeneous steady states:

\begin{eqnarray}
\partial_t u =  F(u)  &+& \partial^2_x u  \label{Eq.u.nlc_3_zero.diff} \\
&-& \sigma \left( \int_{-\infty}^{\infty} \mathrm{d}x' G(x') u(x-x') - u(x) \right), \nonumber    
\end{eqnarray}

Note that this type of distributed nonlocal feedback can be seen as a generalization of the discrete nonlocal feedback used, for example, in \cite{DAH08,SCH09c}.

\begin{figure}[t]
\centering
  \resizebox{0.4\textwidth}{!}{
\includegraphics[]{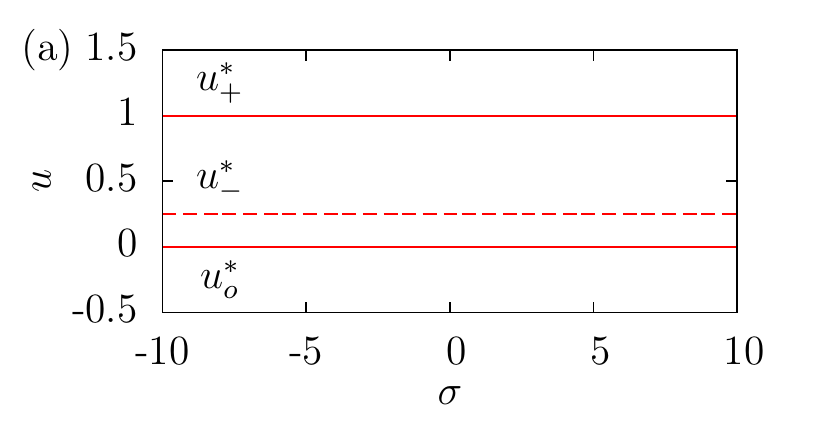}
  }\\
  \resizebox{0.4\textwidth}{!}{
\includegraphics[]{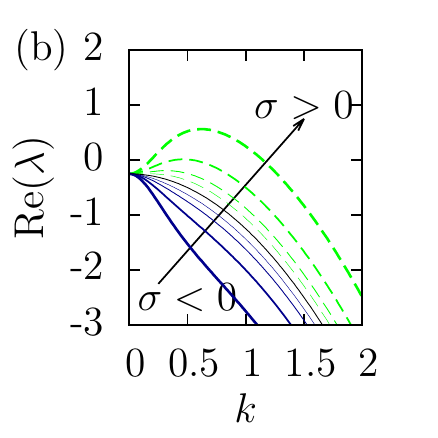}
    \hspace{-1cm}
\includegraphics[]{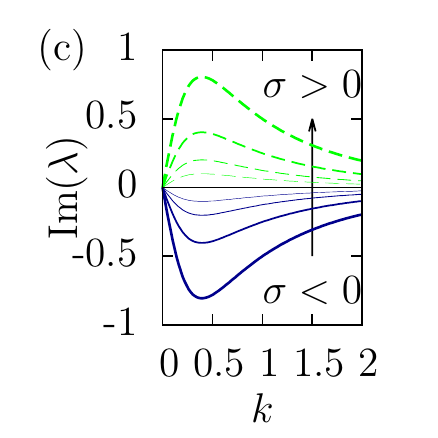}
  }
  
\caption{(a) Spatially homogeneous solutions for various values of $\sigma$. 
Solid and dashed lines represent stable and unstable solutions, respectively. (b),(c) Dispersion relation (Eq.(\ref{Eq.DisRel.w3})) Re $\lambda(k)$, Im $\lambda(k)$ of the solution $u^*_o$ for different values of $\sigma=-2,-1,-0.5,-0.25,0,0.25,0.5,1,2$.
Arrows show the direction of increasing value of $\sigma$ 
(solid blue: $\sigma < 0$, dashed green: $\sigma > 0$).
Parameters $ \alpha=0.25 $, $f=1$, $D=1$, $\xi=2$.} 
\label{figdis}
\end{figure}

\subsection{Stability analysis}

The homogeneous steady state solutions of Eq.(\ref{Eq.u.nlc_3_zero.diff}) are the same as for the reaction-diffusion system without coupling ($u^*_0$, $u^*_+=1$ and $u^*_-=\alpha$) and they are independent of the control parameter $\sigma$, see Fig.\ref{figdis}(a).
The stability analysis of the solution $u^*_0$ gives the dispersion relation for the kernel defined by Eq.(\ref{Eq.ker}):

\begin{eqnarray}
\lambda &=&  F'(u^*_0)- k^2  - \sigma \left( \frac{1}{f + i k \xi + D k^2} - 1 \right). \label{Eq.DisRel.w3}   
\end{eqnarray}

The stability of the homogeneous solution is not changed by the coupling, and the value of the dispersion relation in Eq.(\ref{Eq.DisRel.w3}) is always the same at $k=0$, irrespectively of $\sigma$, see Fig.\ref{figdis}(b),(c). 
The other modes, however, change with the control parameter $\sigma$, and for large positive values of $\sigma$ the system becomes unstable at a characteristic wavenumber $k$, giving rise to a wave instability, see Fig.\ref{figdis}(b),(c).

In the limit of pure diffusion in the fast inhibitor equation (symmetric kernel defined by Eq.(\ref{Eq.diff})), we obtain the following dispersion relation:
\begin{eqnarray}
\lambda &=&  F'(u_0^*)  - k^2  - \sigma \frac{f+Dk^2-1}{f+Dk^2}, \label{Eq.DispRel.diff}   
\end{eqnarray}
which is real; hence only static patterns are expected.

In the limit of pure advection in the fast inhibitor equation (asymmetric kernel defined by Eq.(\ref{Eq.adv})), we obtain the following dispersion relation:
\begin{eqnarray}
\lambda &=&  F'(u_0^*)  - k^2  - \sigma \frac{f + i \xi k - 1}{f + i \xi k}. \label{Eq.DispRel.adv}   
\end{eqnarray}

\subsection{Numerical simulations}
\begin{figure}[t]
 \centering
 \includegraphics[width=0.4\textwidth]{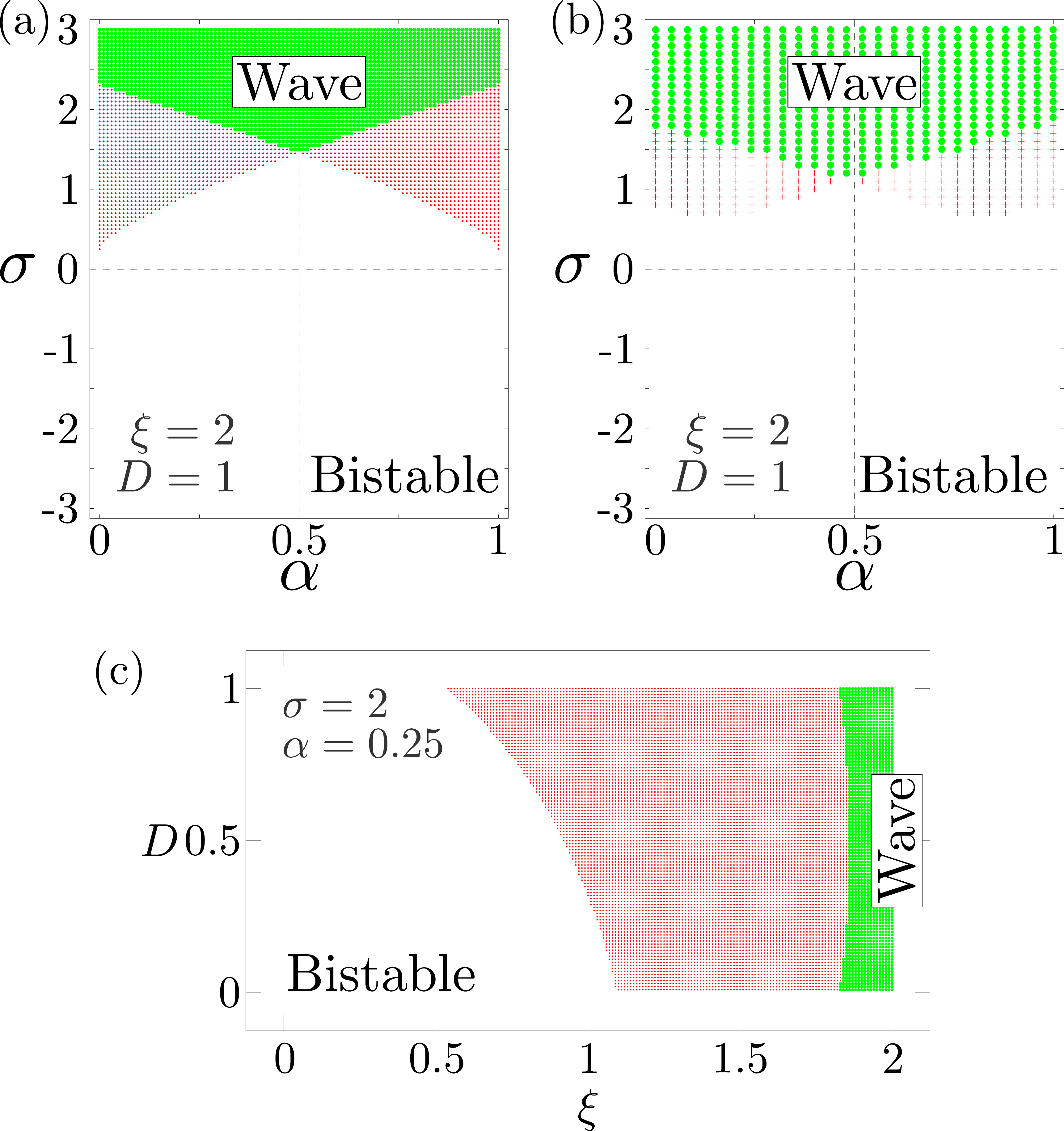}
 \caption{(a) Stability diagram of the homogeneous steady states in the $(\alpha$, $\sigma)$ plane numerically obtained from the dispersion relation Eq.(\ref{Eq.DisRel.w3}). 
 Bistability (white): the two homogeneous steady states are stable.
 Wave instability (light red dots): one of the solutions exhibits a wave instability.
 Strong wave regime (dark green dots): both homogeneous solutions exhibit a wave instability.
 Parameters $f=1$, $D=1$, $\xi = 2$.\\
 (b) Patterns obtained from simulation of Eq.(\ref{Eq.u.nlc_3_zero.diff}). Bistability (white): acceleration or deceleration of the fronts (Fig.\ref{stdiag.accel.suppr}(b)). Wave instability (green circles): travelling wave patterns; (red crosses) coexistence of wave patterns and homogeneous steady state.\\
 (c) Stability diagram of the homogeneous steady states in the $(\xi$, $D)$ plane. Parameters $\sigma = 2$, $\alpha = 0.25$, $f=1$.
 }
 \label{figphasediag2}
\end{figure}

From the dispersion relation Eq.(\ref{Eq.DisRel.w3}) we numerically compute the phase diagrams in Fig.\ref{figphasediag2}(a),(c) representing the stability of the homogeneous steady states $(u^*_o,u^*_+)$. We see that for positive values of $\sigma$ both steady states can change their stability. 

We also perform a series of numerical simulations for different parameter values (Eq.(\ref{Eq.u.nlc_3_zero.diff}) with simulation parameters as in Fig.\ref{fig_stdiag}. The dynamics observed is shown in the phase diagrams of Fig.\ref{figphasediag2}(b). For negative values of the coupling strength $\sigma<0$ the front propagates with modified velocity. However, for sufficiently strong positive coupling $\sigma>0$ waves are observed (green dots and red crosses in Fig.\ref{figphasediag2}(b). The critical value of $\sigma$ for the transition to wave patterns depends on the other parameters. 

In these simulations we observe that the system can exhibit coexistence of homogeneous steady state solutions and travelling waves such as those observed in Fig.\ref{stdiag2}(c),(d). These patterns are denoted by red crosses in Fig.\ref{figphasediag2}(b). Note that travelling wave patterns may occur outside the regime of wave instability of the homogeneous steady states, if only one homogeneous steady state is unstable but also if both of them are stable (see Fig.\ref{figphasediag2}(b) with $\alpha = 0.5$ and $1>\sigma>1.5$).

\section{Discussion}\label{sect.disc}
We have shown that systems of two or three reaction-diffusion-advection equations can be reduced, to a single equation with nonlocal feedback coupling provided that two variables exhibit fast dynamics.
Specifically, we consider FitzHugh-Nagumo type activator-inhibitor systems with a bistable activator dynamics affected by one or two fast inhibitors.
We have shown how different coupling kernels affect the front propagation and generate new spatially periodic patterns.
For the special case of zero diffusion of the second inhibitor, the steady states remain unchanged by the coupling.

Advection in the original system produces asymmetric nonlocal coupling in the reduced system.
Acceleration, deceleration, or suppression of propagating fronts, and 
wave instabilities can be induced by asymmetric nonlocal coupling, 
\emph{i.e.} by advection of a fast inhibitor.

The specific choice of the nonlocal coupling terms, \emph{i.e.} the parameters controlling the fast inhibitor dynamics in the original system, allows for deliberate control of the space-time patterns.
While for symmetric coupling Turing patterns appear, for asymmetric coupling waves dominate. The control parameter is the coupling strength and the instabilities appear for sufficiently large values.

Here, we have considered only linear inhibitor equations, in addition to the first nonlinear activator equation with the nonlinear function $F(u)$. This permits a simple calculation of the equivalent nonlocal coupling kernel.

In summary, we have found that nonlocal coupling can accelerate or slow down propagating fronts, and furthermore symmetric or asymmetric nonlocal coupling can induce the formation of Turing or wave patterns, respectively. The exponential kernels employed here can be derived from extended reaction-diffusion-advection equations in the limit of fast inhibitor dynamics.

\section*{Acknowledgment}
We acknowledge financial support from the German Science Foundation DFG in the framework of SFB 910 ``Control of self-organizing nonlinear systems.''

\bibliography{SiebertAlonso}

\end{document}